\documentclass[twoside]{ilcws10}
\usepackage[latin1]{inputenc}
\usepackage[dvips]{graphicx,epsfig,color}
\usepackage{wrapfig,rotating}
\usepackage{amssymb,amsmath,array}

\begin{document}
\title{
%%%%   Paper title goes here  %%%%%%%%%%%%%%
R\&D Status and Plan for FPCCD VTX
}
%% 
%***********************************************************************
% AUTHORS INFORMATION AREA
%***********************************************************************
\author{Yasuhiro Sugimoto$^1$, 
Hirokazu Ikeda$^2$,
Kennosuke~Itagaki$^3$,
Akiya Miyamoto$^1$, \\
Yousuke~Takubo$^3$,
and Hitoshi~Yamamoto$^3$
% Optional short acknowledgment: remove next line if non-needed
%\thanks{On behalf of the FPCCD VTX Collaboration}
% DO NOT MODIFY THE FOLLOWING '\vspace' ARGUMENT
\vspace{.3cm}\\
% Addresses and institutions (remove "1- " in case of a single institution)
1- KEK, High Energy Accelerator Research Organization \\
Tsukuba, Ibaraki 305-0801, Japan
\vspace{.1cm} \\
2- JAXA, Japan Aerospace Exploration Agency \\
Sagamihara, Kanagawa 229-8510, Japan
\vspace{.1cm} \\
3- Department of Physics, Tohoku University \\
Sendai 980-8578, Japan
}
%%***********************************************************************
% END OF AUTHORS INFORMATION AREA
%***********************************************************************

\maketitle

\begin{abstract}
Fine pixel CCD (FPCCD) is an option for the sensor
used for the ILC vertex detector to reduce the pixel
occupancy due to the high background rate 
near the interaction point. In this paper, we report
on the R\&D status of FPCCD sensors and the R\&D plan
for the sensors and the cooling system.
\end{abstract}

\section{Introduction}
Fine pixel CCD (FPCCD) is one of the sensor candidates
for the vertex detector at ILC experiment~\cite{ildloi}.
FPCCD vertex detector was proposed~\cite{sugimoto05} 
as an option compatible with the beam background
condition of super-conducting RF technology of the 
linear collider.
The FPCCD sensors have the pixel size 
as small as $\sim 5$~$\mu{\rm m}$.
The sensitive layer is as thin as $\sim 15$~$\mu{\rm m}$
and fully depleted to suppress the number of hit pixels 
due to charge diffusion. Because of the huge number of 
the pixels, the pixel occupancy due to pair background
can be kept low even if the signal is accumulated for
one train of the ILC beam collision. In addition,
background rejection can be achieved by the hit cluster
shape analysis thanks to
the fine pixel size smaller than the sensitive layer thickness.
In order to read out the huge number of pixels in the
beam-train interval of 200~ms, multi port readout 
of the FPCCD sensors is indispensable. 
The horizontal
shift registers lie along the beam direction and
embedded in the image area. 
Therefore the number
of horizontal (serial) shift is much more than the number
of vertical (parallel) shift. This layout gives better
radiation immunity than the inverse layout
(long parallel shift and short serial shift)
like SLD vertex detector.

Our R\&D activity for the FPCCD vertex detector has started
in 2006. It includes R\&D for FPCCD sensors, 
readout ASICs~\cite{takubo08, itagaki},
peripheral electronics such as clock gate drivers,
wafer thinning and low mass ladder, cooling system,
and simulation study for background rejection~\cite{yoshida}.
In this paper, R\&D status of the sensor is described in
section~\ref{sensor}, and R\&D plan for the cooling system
is described in section~\ref{cooling}.

\section{Sensor R\&D}
\label{sensor}

We have developed several types of prototype sensors
collaborating with Hamamatsu Photonics.
As the first step of the FPCCD sensor R\&D, we have developed
fully depleted CCDs with a standard pixel size of $24\ \mu {\rm m}$.
It has been demonstrated that the epitaxial layer of these CCDs is 
fully depleted by using high resistive epitaxial layer 
of $24\ \mu {\rm m}$ thickness~\cite{sugimoto07}.

Using the same type of the wafer, we have made a FPCCD prototype
with the pixel size of $12\ \mu {\rm m}$~\cite{sugimoto08}. 
This prototype has four output nodes and four horizontal shift registers.
The horizontal registers are embedded in the image area and 
sensitive to light, as well as to charged particles.

\begin{figure}
\centerline{\includegraphics[width=0.6\columnwidth]{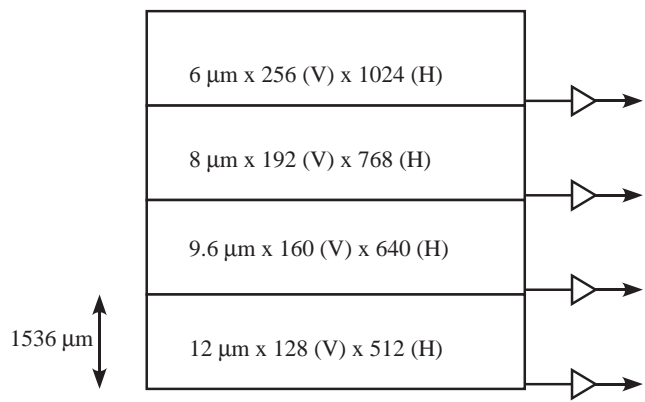}}
\caption{Pixel configuration of a prototype FPCCD sensor in FY2009}
\label{Fig:pixconf}
%\end{wrapfigure}
\end{figure}

\begin{figure}
\centerline{
\includegraphics[width=0.45\columnwidth]{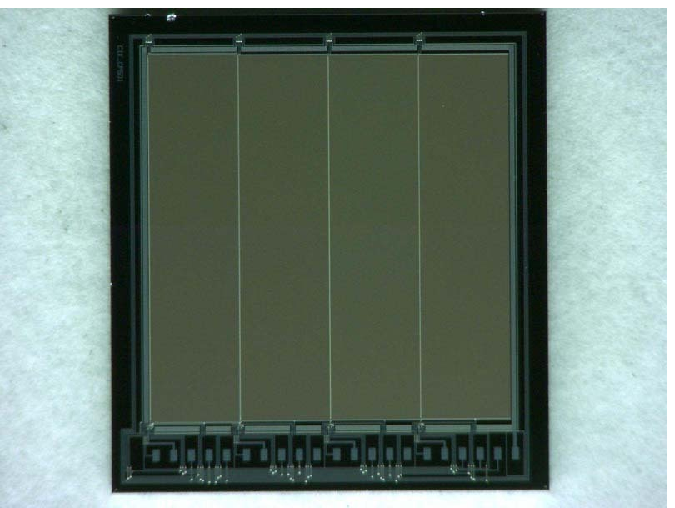}
\hspace{2em}
\includegraphics[width=0.45\columnwidth]{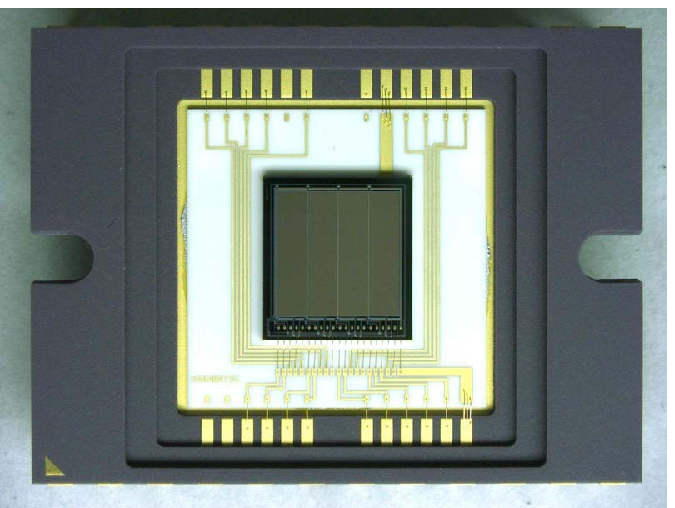}
}
\caption{Bare chip (left) and packaged chip (right) of 
a prototype FPCCD sensor in FY2009}
\label{Fig:ccdpict}
\end{figure}

In FY2009, we have fabricated a new prototype of 
FPCCD sensors with smaller pixel size.
Figure~\ref{Fig:pixconf} shows pixel configuration
of the new prototype sensor. It has four types of
pixel size, 12, 9.6, 8, and 6~$\mu$m. 
The smallest pixel size of 
$6\ \mu {\rm m}$ square is very close to our R\&D goal. 
The chip size (sensitive area) is 6.1~mm square.
Pictures of a bare chip and a
packaged chip are shown in Figure~\ref{Fig:ccdpict}.
Bright lines between four areas are horizontal
shift registers. Because these horizontal registers
are also sensitive, there is no sensitivity gap
between four areas. 

%\begin{wrapfigure}{r}{0.4\columnwidth}

These prototype chips have just been delivered 
at the end of March 2010. Detailed study of these 
chips will be done in FY2010.
We would 
study on  S/N ratio, incident angle measurement
using the hit cluster shape, spatial resolution, two-track
separation, and radiation immunity.

\section{Cooling system}
\label{cooling}

The estimated power consumption of the FPCCD vertex detector is 
about 80~W including the sensors and the readout front-end ASICs 
inside a cryostat. Additional power for peripheral circuits
such as clock drivers is also dissipated, but these circuits
would be placed outside the cryostat of the vertex detector.
The FPCCD vertex detector is expected to be operated
at low  temperature of $\sim -40^\circ$C inside the cryostat
in order to minimize the effect of radiation damage.

A possible cooling system for the FPCCD vertex detector
is flow of cool nitrogen gas. In order to remove
heat generated by 80~W power, a flow rate of 
3~l/s is necessary with $\Delta T=20^\circ$C.
By using 1~cm diameter gas tube, the gas velocity
is as high as 40~m/s. This high-speed gas flow
could cause vibration of the ladder. In order to
mitigate this risk, we started considering 
liquid cooling using two-phase CO$_2$.
Because ladders of the FPCCD vertex detector
have the major heat source at both ends
(output amplifiers of the CCDs and the ASICs), 
cooling with liquid coolant removing the heat
from the ladder ends could be a solution.

Two-phase CO$_2$ cooling uses latent heat of 
evaporative liquid CO$_2$. With two-phase
liquid cooling, incoming heat
at the evaporator ({\it ie.} detector) is used
for evaporating the liquid rather than raising the
temperature. Therefore we can expect almost 
isothermal cooling along the cooling tube.

Compared to other two-phase coolant such as
perfluorocabon (C$_n$F$_{2n+2}$), which  is
used for example in ATLAS inner detector~\cite{attree08},
liquid CO$_2$ has larger latent heat as shown in
Table~\ref{tab:coolant}.  
If we allow 50\% evaporation of the liquid CO$_2$, 
a flow rate of only 0.5~g/s is necessary.
Since CO$_2$ is circulated under higher pressure
(1~MPa at $-40^\circ$C and 5~MPa at $15^\circ$C),
volume of the vapor remains smaller than
perfluorocarbon.
For these reasons, we can use thinner tube
for the two-phase CO$_2$ cooling.

In principle,
two-phase cooling can be used 
in the temperature range
between the triple point and the critical point.
Since our target temperature of $-40^\circ$C is
close to the triple point of CO$_2$, 
it may be challenging to achieve this goal.
 
%\begin{wraptable}{l}{0.5\columnwidth}
\begin{table}
\centerline{\begin{tabular}{|l|r|r|r|}
\hline
                             & CO$_2$  & C$_2$F$_6$  & C$_3$F$_8$  \\
\hline
Latent heat at $-40^\circ$C (J/g)  & $321$ & $\sim 100$ & $\sim 110$ \\
\hline
Triple point ($^\circ$C)     & $-56.4$ & $-97.2$ & $-160$ \\
\hline
Critical point ($^\circ$C)   & $31.1$  & $19.7$  & $71.9$ \\ 
\hline
\end{tabular}}
\caption{Properties of coolant.}
\label{tab:coolant}
\end{table}
%\end{wraptable}

In the past,
two-phase CO$_2$ cooling  has been used for
AMS tracker~\cite{delil03} and LHCb-VELO~\cite{beuzekom07}.
It is also proposed for ILD TPC cooling by NIKHEF group.
Our group is trying to organize an R\&D collaboration
for two-phase CO$_2$ cooling together with ILD TPC group,
Belle-II vertex group, and KEK cryogenic group in Japan.

\section{Summary}

We have developed fully depleted CCDs with standard (24~$\mu$m),
medium (12~$\mu$m), and finally fine (6~$\mu$m) pixel size
for the FPCCD vertex detector. Detailed study on the FPCCD
prototype sensors will be done in FY2010. 
Cooling system using two-phase CO$_2$ is an interesting option
for the FPCCD vertex detector because the FPCCD vertex detector has 
main heat source at the ladder ends. We will start R\&D on
two-phase CO$_2$ cooling system collaborating with
ILD TPC group, Belle-II vertex group, and KEK cryogenic
group in Japan.

\section*{Acknowledgements}
This work is partly supported by Creative Scientific 
Research Grant No.18GS0202 of Japan Society for Promotion
of Science (JSPS).

%\section{Bibliography}
%
%If possible please use the bibtex information as given by SPIRES
%to make the citations~\cite{parton_qed} uniform and follow the 
%examples~\cite{parton_qed,H1,DVCS,pomeron} given below.
%Note that there is a (non-breaking) space before \verb?\cite?.
%
% ****************************************************************************
% BIBLIOGRAPHY AREA
% ****************************************************************************

\begin{footnotesize}
% IF YOU DO NOT USE BIBTEX, USE THE FOLLOWING SAMPLE SCHEME FOR THE REFERENCES
% ----------------------------------------------------------------------------

% ----------------------------------------------------------------------------

\end{footnotesize}

% ****************************************************************************
% END OF BIBLIOGRAPHY AREA
% ****************************************************************************


\begin{thebibliography}{99}
% Please replace the numbers for   contribId   and   sessionId
% in the following URL. You can get this information by going to 
% http://indico.cern.ch/confAuthorIndex.py?confId=2628
% and search for your contribution and click on the title
% Be aware: '&amp;' must be replaced by simple '&' as in example below
%------- replace following references ;-)
\bibitem{ildloi} T. Abe{\it et~al.}, ILD Letter of Intent, KEK Report 2009-6. 
\bibitem{sugimoto05} Y. Sugimoto, Proceedings of International 
Linear Collider Workshop LCWS05, Stanford, CA, March 2005, pp.550-554.
\bibitem{takubo08} Y. Takubo {\it et~al.}, arXiv:0901.3427[physics.ins-det] (2009).
\bibitem{itagaki} K. Itagaki {\it et~al.}, in these proceedings.
\bibitem{yoshida} K. Yoshida {\it et~al.}, in these proceedings.
\bibitem{sugimoto07} Y. Sugimoto, Proceedings of International 
Linear Collider Workshop LCWS2007, DESY, Hamburg,May 2007, 483-485.
\bibitem{sugimoto08} Y. Sugimoto {\it et~al.}, arXiv:0902.2067[physics.ins-det] (2009).
\bibitem{attree08} D. Attree  {\it et~al.}, JINST 3:P07003 (2008).
\bibitem{delil03} A.A.M. Delil {\it et~al.}, NLR-TP-2003-001 (2003).
\bibitem{beuzekom07} M. Van Beuzekom {\it et~al.}, PoS(Vertex 2007)009 (2007).


\end{thebibliography}
\end{document}